\documentclass[aps,pre,reprint]{revtex4-1}
\usepackage{amsmath}
\DeclareMathOperator*{\argmax}{arg\,max}
\usepackage[latin1]{inputenc}
\usepackage{graphicx}

\usepackage[]{hyperref}
\usepackage{bm}

\usepackage{epsfig}
\usepackage{float}

\usepackage{color}
\usepackage{xcolor}
\hypersetup{
    colorlinks,
    linkcolor={red!50!black},
    citecolor={blue!50!black},
    urlcolor={blue!80!black}
}

\begin{document}
\title{The excluded area of two-dimensional hard particles}

\author{Thomas Geigenfeind}
\affiliation{Theoretische Physik II, Physikalisches Institut,
 Universit{\"a}t Bayreuth, D-95440 Bayreuth, Germany}

\author{Daniel de las Heras}
\email{delasheras.daniel@gmail.com}
\homepage{www.danieldelasheras.com}

\affiliation{Theoretische Physik II, Physikalisches Institut,
 Universit{\"a}t Bayreuth, D-95440 Bayreuth, Germany}

\date{\today}
\begin{abstract}
The excluded area between a pair of two-dimensional hard particles with given relative orientation 
is the region in which one particle cannot be located due to the presence of the other
particle. The magnitude of the excluded area as a function of the relative particle orientation
plays a major role in the determination
of the bulk phase behaviour of hard particles. We use principal component analysis to identify the
different types of excluded area corresponding to randomly generated two-dimensional hard particles
modeled as non-self-intersecting polygons and star lines (line segments radiating from a common origin). Only 
three principal components are required to have an excellent representation of the value of the excluded 
area as a function of the relative particle orientation. Independently of the particle shape, the minimum value of the excluded area is always achieved 
when the particles are antiparallel to each other. The property that affects the value of the excluded area most strongly is the
elongation of the particle shape. Principal component analysis identifies four limiting cases of excluded areas 
with one to four global minima at equispaced relative orientations.
We study selected particle shapes using Monte Carlo simulations.
\end{abstract}

\maketitle

\section{Introduction}

Hard body models, for which the interaction potential is infinite if two particles overlap and zero otherwise,
are excellent candidates to model colloidal particles, which are dominated by excluded volume.
Hard body models are also relevant to understand how the microscopic properties at the particle level determine the macroscopic
properties of the system such as its bulk phase behaviour. Since the then unexpected classical result of fluid-solid phase transition
in a system of hard spheres~\cite{doi:10.1063/1.1743957}, the bulk phase behaviour of several three-dimensional
hard body models has been analysed by both computer simulations and theoretical approaches such as density functional theory.
Anisotropic hard particles form a surprisingly
rich variety of mesophases such as uniaxial, biaxial and cubatic nematics, as well as cholesteric,
smectic, and columnar phases. We refer the reader to Ref.~\cite{Mederos2014} for a recent review.

The phase behaviour of several two-dimensional hard models has been also reported in the literature. Examples are 
hard disks~\cite{PhysRevE.55.4245,PhysRevLett.118.158001,doi:10.1063/1.3687921}, needles \cite{Frenkel1985}, rectangles~\cite{doi:10.1063/1.1849159,Donev2006,thomas}, discorectangles \cite{Bates2000,PhysRevE.62.5081,doi:10.1063/1.1849159}, 
triangles \cite{Gantapara2015, Martinez-Raton2018}, squares \cite{WOJCIECHOWSKI2004},
rounded squares \cite{Avendano2012}, pentagons \cite{Schilling2005}, hexagons~\cite{doi:10.1021/nl4046069,PhysRevMaterials.3.015601}, ellipses \cite{Cuesta1990}, zigzags \cite{Varga2009d},
hockey sticks \cite{Martinez-Gonzalez2013}, banana-like \cite{Martinez-Gonzlez2012}, and allophiles \cite{Harper2015}.

Phase transitions in hard bodies are driven by entropy. At sufficiently low density,
the entropy of the ideal gas dominates, and the system remains
isotropic with neither positional nor orientational order. As the density increases, excluded volume effects become more important
and the gain in configurational entropy can drive a transition to a phase with orientational and/or positional order. 

The study of excluded volume effects in hard bodies starts with the properties of the excluded volume between two particles, which
plays a role similar to the pair interaction potential in soft systems (i.e., with continuous interaction potentials).
Here, we restrict ourselves to two-dimensional particles. 
As a result of the potential being infinite if two particles overlap, there exists around each particle 
an exclusion region in which no other particle can be located.
The phase behaviour of the system is determined by the properties of this complicated
many-body exclusion region, which depends on the positions and orientations of all particles
in the system. Within a mean field-like approach the properties of the many-body exclusion region
are fully characterized by the excluded area, which is the area inaccessible to one particle due to the presence
of another particle. The excluded area alone does not determine the complete phase behaviour of the system but
it plays a vital role in the determination of the type of stable phases. The value of the excluded area, $A_{\text{exc}}$, 
depends on the particle shape. For a given particle shape $A_{\text{exc}}$ is a function of the relative orientation between the two
particles, $\phi$. For example, the value
of the excluded area between two rectangles is minimum if their relative angle $\phi$ is either $0$ (parallel) or $\pi$ (antiparallel).
For squares, the value of the excluded area is minimum if $\phi=0,\pi/2,\pi$, and $3\pi/2$. Although different particle shapes might 
generate different excluded areas, it is evident that not any function $A_{\text{exc}}(\phi)$ corresponds to a valid particle shape 
(e.g., the value of the excluded area cannot be negative). We investigate here the admissible shapes of the function $A_{\text{exc}}(\phi)$. To this end, 
we apply principal component analysis~\cite{PearsonK.1901,Wold1987} (PCA) to a set of excluded areas corresponding to randomly generated two-dimensional hard bodies.
PCA is an unsupervised learning algorithm intended to simplify the complexity of a high-dimensional data set. Mathematically, PCA
is an orthogonal transformation of the data to a new basis in which the basis vectors are sequentially chosen such that the variance of the projection
of the original data onto the new basis vectors is as large as possible. As a result of our PCA we identify four special or limiting types of excluded
area with one to four global minima as a function of the relative orientation.
Finally, we perform Monte Carlo simulations for selected particle shapes.

\section{Methods}

We aim at finding the relevant types of excluded areas $A_{\text{exc}}(\phi)$ in two dimensions. To this 
end we first generate random hard particles, and then apply principal component analysis to
the excluded area between two identical particles.

\subsection{Particle generation}

\begin{figure*}
\includegraphics[width=0.8\textwidth,angle=0]{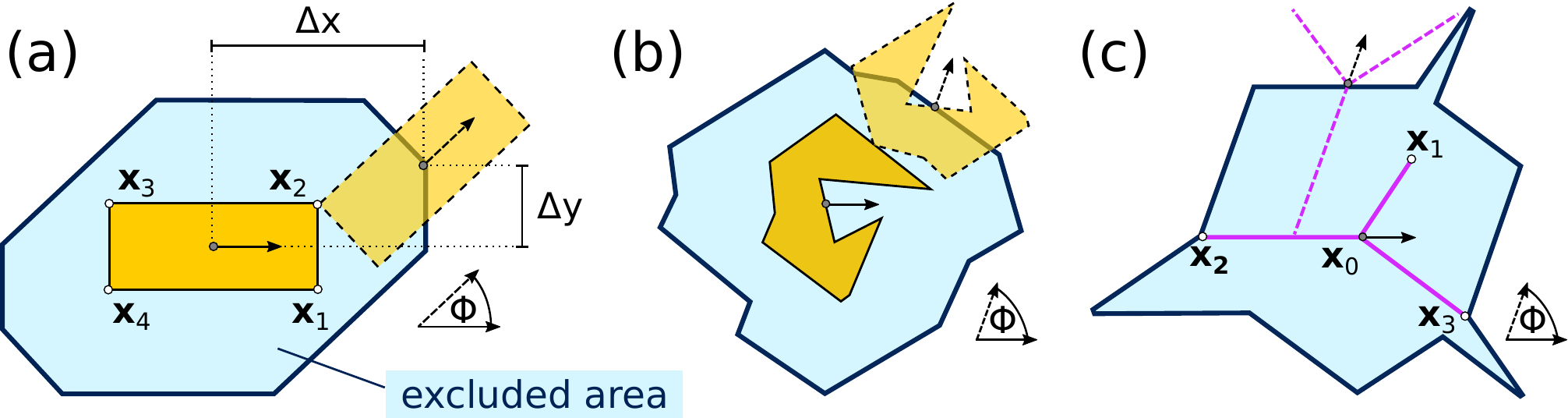}
\caption{Excluded area. (a) A rectangle (solid orange) with vertices (empty circles) at $\mathbf{x}_i$, $i=1,2,3,4$ and a
second identical rectangle (dashed orange) shifted by $(\Delta x,\Delta y)$ and rotated by $\phi$ with respect to
the first particle. The blue area is a graphical representation of the excluded area for the center of the second rectangle at this relative orientation.
The black arrow indicates the orientation of the particles (with respect to an arbitrary axis).
(b) Graphical representation of the excluded area (at a given relative orientation) of a non-self-intersecting polygon with $10$ randomly generated vertices.
(c) Excluded area of a randomly generated star line (magenta) with three vertices $\mathbf{x}_i$ connected to the point $\mathbf{x}_0$.}
\label{fig1}
\end{figure*}

We generate two-dimensional hard particles following either of two procedures. 
In the first one, each particle is modeled as a simple (non-self-intersecting) polygon, as sketched in panels (a) and
(b) of Fig.~\ref{fig1}. A polygon is defined by a closed and ordered set of $m$ random vertices
connected via straight edges. For each polygon the coordinates of each vertex $\mathbf{x}_i$, $i=1,...,m$
are obtained by sampling from uniform distributions a random radius $r_i \in (0,1]$ and a random angle $\phi_i  \in [0,2\pi]$.
We allow both convex and concave shapes, but restrict the set to simple, non-self-intersecting, polygons,
i.e., the line segments connecting the vertices of a polygon are not allowed to intersect (apart from the end points of
two neighbouring edges that are joint at one vertex). Furthermore, no two vertices have the exact same coordinates.
If the random creation of vertices of a polygon leads to self-intersections,
we perform lin 2-opt moves until all intersections are removed \cite{Lin1965}. The algorithm works as follows:
Assume the line segment connecting vertices $\mathbf{x}_{i-1}$ and $\mathbf{x}_{i}$
and the line segment connecting vertices $\mathbf{x}_{j-1}$ and $\mathbf{x}_{j}$ intersect ($j-i>1$ and
periodic vertices). Then, we reverse the order of all vertices between $i$ and $j$, i.e. the sequence
$\{i-1,i,..,j-1,j\}$ becomes $\{i-1,j-1,..,i,j\}$. We perform as many lin 2-opt moves as required
to remove all self-intersections. 

To increase the spectrum of particle shapes we also generate particles modeled as star lines, see Fig.~\ref{fig1}c.
A star line has a center $\mathbf{x}_0$ connected via line segments to $m$ vertices located at $\mathbf{x}_i$, $i=1,...,m$.
The simple polygons contain the star lines as a limiting case of polygons with zero area. However, it is very unlikely to
generate a star line with the method we use to generate polygons. 

As a final step, each particle (polygons and star lines) is rescaled to the maximum possible size that fits in a square bounding
box of length $h=1$, which defines our unit of length.

\subsection{Computation of the excluded area}

In what follows and for simplicity we refer to $A_{\text{exc}}(\phi)$ as simply the excluded area. 
To calculate the excluded area between two particles (let them be polygons or star lines) at a given relative orientation $\phi$
we use the fact that whenever two identical particles overlap then at least two edges intersect.
We first fix the relative orientation between two particles $\phi$, and the position of particle $1$ (at the origin). 

Next, we fix the $y$-distance between the two particles $\Delta y$ and calculate the interval(s) in $x$ for which
the two particles overlap. We loop over all pairs of edges between particle $1$ and $2$ checking
for which $x$-distances between the particles $\Delta x$ the edges overlap. The distance $\Delta x$ ($\Delta y$) is
calculated as the separation in $\hat {\bf x}$ ($\hat {\bf y}$) between two reference points located in the particles such as e.g. the centers of masses.
For a fixed value of $\Delta y$, a pair of edges intersects either not at all or in at most one connected interval of
$\Delta x$. The bounds of this interval represent the cases for which a vertex
of one edge lies on top of the other edge. Hence, the interval in $\Delta x$ for which two edges
overlap can be obtained by checking the values of $\Delta x$ for which each of the four involved vertices lies on top of the other
edge. Combining the overlapping intervals of all possible pairs of edges gives the complete overlap  in
$x$-direction for a given value of $\Delta y$.

Next, we move particle $2$ in y-direction in discrete steps of size $\Delta/h\approx5\cdot10^{-4}$ between $\Delta y_{\text{min}}(\phi)$
and $\Delta y_{\text{max}}(\phi)$ which are the minimum and maximum values of $\Delta y$ for which
overlap is possible, respectively. $\Delta y_{\text{min}}$ ($\Delta y_{\text{max}}$) occurs when the vertex with the highest (lowest)
$y$-coordinate of particle $2$ and that with the lowest (highest) $y$-coordinate of particle $1$ have the same $y$-coordinate.

Integrating the overlapping intervals between  $\Delta y_{\text{min}}$ and  $\Delta y_{\text{max}}$ yields
the the excluded area for the selected orientation. Finally, we repeat the process for $n_\phi=360$ relative orientations $\phi\in[0,2\pi]$
between the particles and normalize the excluded area according to
\begin{equation} \label{eq_norm}
\frac{1}{2\pi h^2}\int_{0}^{2\pi}A_{\text{exc}}(\phi) d\phi=1.
\end{equation}

\subsection{Principal Component Analysis}
We apply principal component analysis~\cite{PearsonK.1901,Wold1987} to the excluded areas $A_{\text{exc}}(\phi)$ generated by the above method.
All data is organized in a data matrix $\mathbf{X}$. Each row contains the excluded area as a function of the relative orientation
for one randomly generated particle.
In each column we store the values of the excluded areas at a given relative orientation for all samples in the system.
First, the data is centered to facilitate the following calculations. That is, from each column of $\mathbf{X}$ the mean is subtracted.
We then apply the PCA algorithm. PCA uses an orthogonal transformation to represent the data in a new orthogonal basis.
In this basis the first basis vector $\mathbf{w}_1$ (first principal axis) is chosen such that the variance of the projection of the data onto
this vector is as large as possible, that is
\begin{equation} \label{eq_PCA}
\mathbf{w}_1= \underset{\Vert\mathbf{w}_1 \Vert = 1 }{\argmax} \left \{ \Vert\mathbf{X}\mathbf{w}_1 \Vert^2 \right \}=\underset{\Vert\mathbf{w}_1 \Vert = 1 }{\argmax} \left \{ \Vert\mathbf{c}_1 \Vert^2 \right \}.
\end{equation}
Here, the $i$-th component of the vector $\mathbf{c_1}$ is the first principal component of the $i-$th sample.
The variance of the following basis vectors, $\mathbf{w}_j$ with $j>1$, is also maximized under the constraint that each vector is orthogonal to the preceding ones.
The new basis vectors are called principal axes and the components of a vector expressed in this basis are the principal components. 
It is often the case that only the first principal components have high variance and are therefore relevant to describe the data. Hence,
PCA allows a meaningful dimension reduction, while retaining as much information as possible.

Mathematically, the principal axes are the eigenvectors of the covariance matrix of the data. The corresponding eigenvalues are 
the variance of the respective principal component~\cite{Joliffe}. For the actual implementation of PCA we use the OpenCV library \cite{Bradski2000}. 

\section{Results}
We start the results section analysing the excluded area of regular polygons. Next we 
show the PCA of the excluded area of randomly generated particles. We end the section
with Monte Carlo simulations of selected particle shapes.
\subsection{Regular Polygons}

\begin{figure} 
\includegraphics[width=0.90\columnwidth,angle=0]{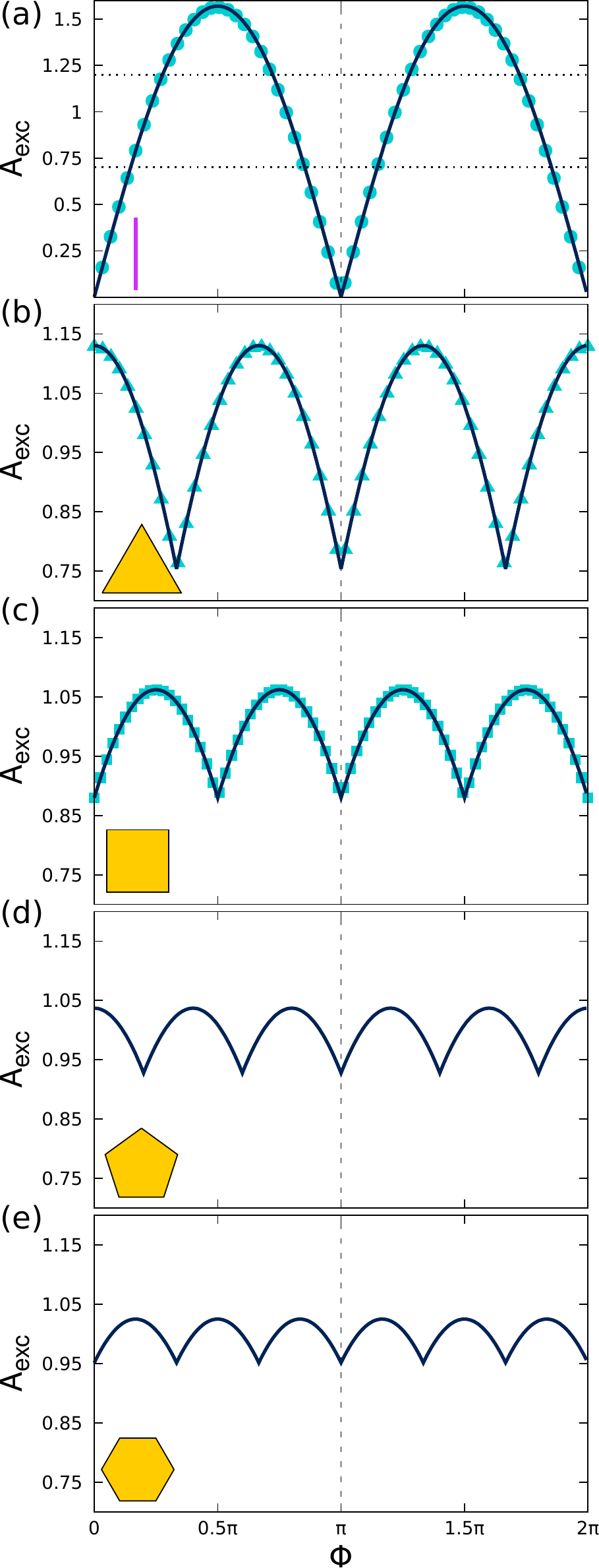}
\caption{Normalized excluded area for regular polygons as a function of the relative orientation $\phi$: (a) line segment,
(b) equilateral triangle, (c) square, (d) pentagon, and (e) hexagon. The insets show the particles. Solid lines are 
numerical results. The symbols are analytic expressions of the excluded area of a line segment (a), an equilateral triangle (b) and a square (c).
The dotted horizontal lines in (a) indicate the vertical range used in panels (b) to (e).
}
\label{fig2}
\end{figure}

The normalized values of the excluded areas for a line segment and for the regular
polygons with $3-6$ vertices are presented in Fig. \ref{fig2}. To check the accuracy of the numerical calculation
of the excluded area we have compared the numerical results against analytic expressions for line segments, equilateral triangles and squares, see
Fig.~\ref{fig2} panels (b) and (c).

For regular polygons the excluded area contains as many minima (and maxima) as
vertices of the polygon since the rotational symmetry of the particle is also present
in the excluded area. The excluded area of the
line segment is a direct extension of this trend as it contains two minima (and two maxima).

All the excluded areas are symmetric with respect to the relative orientation $\phi=\pi$.
This is a general property of the excluded area between any two identical particles that can be easily understood
as follows. The relative orientations $-\phi$ and $\phi$ are degenerate as they correspond to an (irrelevant) swap of the two
identical particles. Hence the excluded area is symmetric with respect to $\phi=0$, which implies also 
the symmetry with respect to $\phi=\pi$. Therefore, although in Fig. \ref{fig2} we present the excluded area of particles
with mirror symmetry, the symmetry of the excluded area around $\phi=\pi$ 
is also present for particles with no spatial symmetries.

As expected, the difference between the maximum and the minimum of the excluded area of 
regular polygons decreases by increasing the number of vertices of the regular polygon.
The limit of a regular polygon with an infinite number of vertices is a disk, for which
the excluded area does not depend on the relative orientation. The difference in
the excluded area for different orientations is correlated with the increase in configurational entropy that occurs
when the particles organize in a state with orientational order. Hence, ordering effects are stronger for particles with more variance
in $A_{\text{exc}}(\phi)$. For example, in Ref.~\cite{Anderson2017a} Anderson et. al. found that regular polygons with more
than $6$ vertices already melt like disks.

\subsection{Principal Component Analysis}
In what follows we show the result of the PCA applied to a set of $9.3\cdot10^4$ randomly generated particles of which
$4.8\cdot10^4$ are polygons and  $4.5\cdot10^4$ are star lines. For the polygons
we generate $6\cdot10^3$ samples for each number of vertices in the interval $[3,10]$. 
An example of a non-self-intersecting polygon with $10$ vertices is shown in Fig.~\ref{fig1}b.
The number of vertices of the star lines is uniformly distributed between $2$ and $10$.
We investigated ten different sets, each one containing the same number of randomly generated particles.
All sets produced the same results up to small numerical inaccuracies.  
 
\begin{figure}
\includegraphics[width=0.9\columnwidth,angle=0]{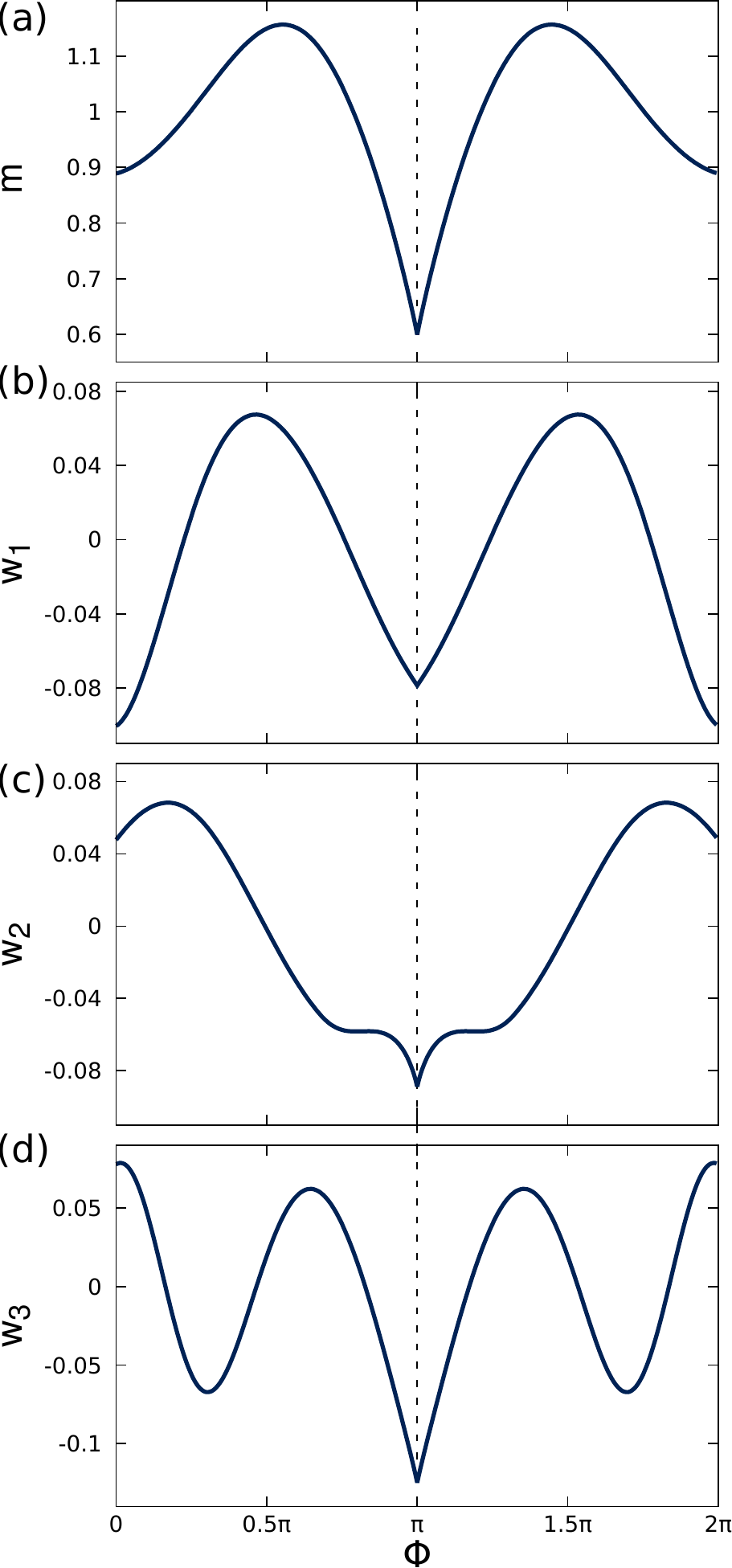}
\caption{Mean excluded area $m$ (a), first $w_1$ (b), second $w_2$ (b), and third $w_3$ (c) principal
axes as a function of the relative orientation between particles $\phi$.}
\label{fig3}
\end{figure}

As described in the preceding section, the first step in PCA consists of centering the
data by removing the column-wise mean of the data matrix.
The mean excluded area $m(\phi)$, see Fig.~\ref{fig3}a, has a global minimum at $\phi=\pi$ and a local minimum at $\phi=0$.
The mean excluded area manifests a common feature of all excluded areas we have calculated:
The excluded area between two identical particles has always the global minimum at $\phi=\pi$, regardless of the
shape of the particles. This intuitive feature has been proven for convex shapes \cite{Palffy-Muhoray2014}. 
Our particles are both convex and concave and we have always observed the global minimum to be located at $\phi=\pi$.

Panels (b) to (d) of Fig. \ref{fig3} show the first three principal axes $w_i(\phi)$, $i=1,2,3$.
The first (b) and third (d) principal axes have qualitatively similar shapes to the excluded area of a 
line segment (Fig.~\ref{fig2}a) and of an equilateral triangle (Fig.~\ref{fig2}b), respectively.
The second principal axis $w_2(\phi)$, Fig.~\ref{fig3}c,
has a pronounced minimum at $\phi=\pi$. Excluded areas with this feature play an important role in the PCA analysis, as we discuss below.

\begin{figure}
\includegraphics[width=1\columnwidth,angle=0]{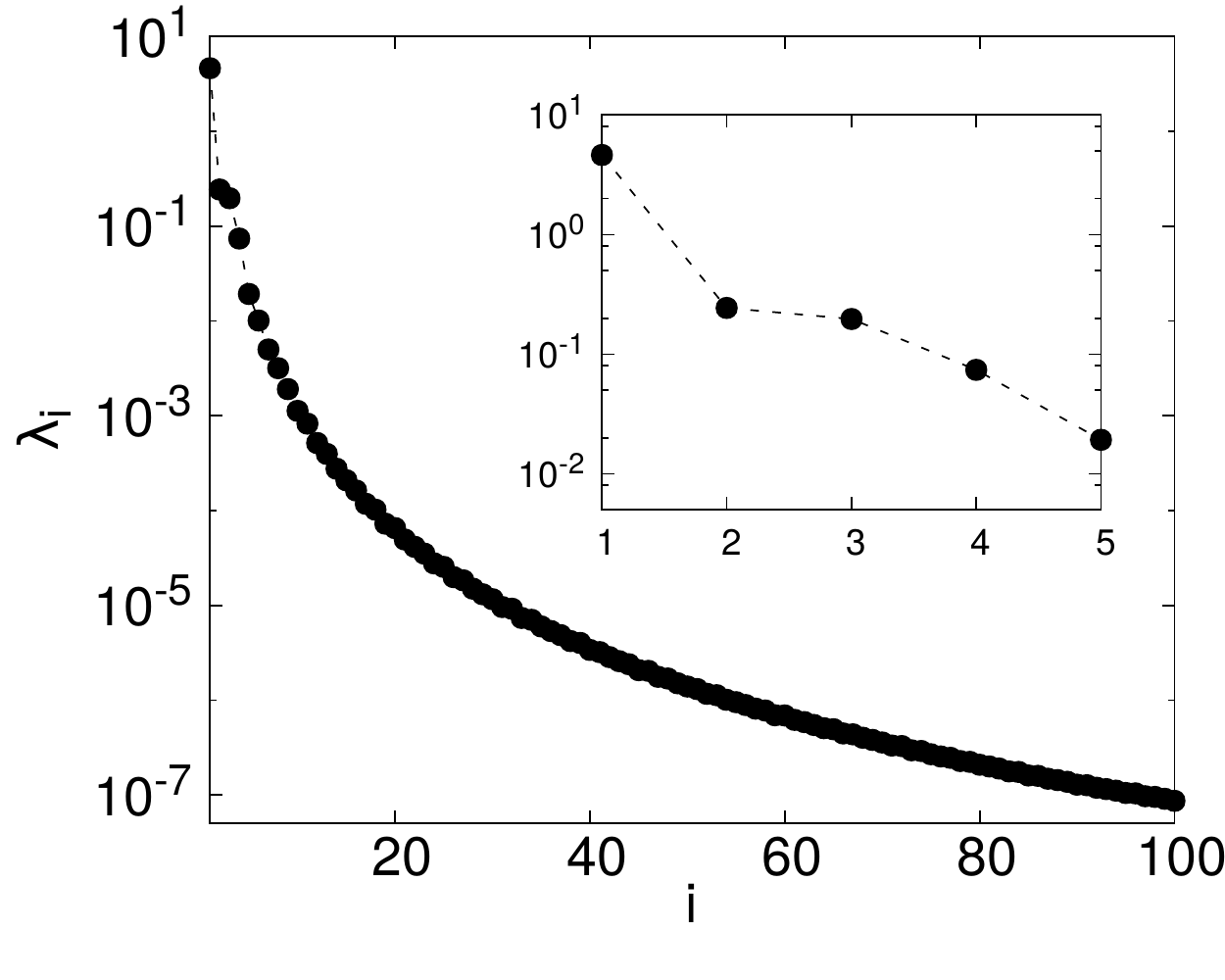}
\caption{Semi-log plot of the first $10^2$ eigenvalues $\lambda_i$ versus the index $i$. The inset is a close view of the first five eigenvalues.}
\label{fig4}
\end{figure}

In Fig. \ref{fig4} a semi-logarithmic plot of the eigenvalues of the first $10^2$ principal axes is presented.
The eigenvalue of a principal axis is 
the variance of the respective principal component~\cite{Joliffe}.
The first and the second eigenvalue differ in one order of magnitude, and higher order eigenvalues decrease
very fast in magnitude.
Due to the rapid decrease of the eigenvalues, Fig.~\ref{fig4}, we achieve an excellent representation of the excluded
area using only the first three principal components. A measure of how well the data are represented using the first
$n$ components is the explained variance $\sigma_n$, which is the sum of the eigenvalues associated with the $n$
first principal components divided by the sum of all eigenvalues. In our case, using three principal components we find
$\sigma_3\approx0.98$ and therefore we are confident that most of the full information is already contained in the first three
components. Using the principal components of a given particle, its approximated excluded area can be reconstructed
by calculating the sum of the first three principal axes (see Fig.~\ref{fig3}b-d) multiplied with their respective principal
components and adding the mean (Fig.~\ref{fig3}a). In other words, the reconstructed excluded area is a linear combination of
the principal axes and the mean (see Fig.~\ref{fig3}), with the principal components being the coefficients of the linear combination.
The average $L_1$ error
\begin{equation}
s_{L_1}=\langle |A_{\text{exc}}(\phi)-A_{\text{exc}}^{\text{rec}}(\phi)|\rangle 
\end{equation}
between the calculated values of the excluded areas $A_{\text{exc}}(\phi)$ and their reconstructions $A_{\text{exc}}^{\text{rec}}(\phi)$ is $0.012$.
Here, the average is taken over all orientations $\phi$ and over all samples. We show examples of the reconstructed excluded area for relevant 
particle shapes below.

\begin{figure*}
\includegraphics[width=2\columnwidth,angle=0]{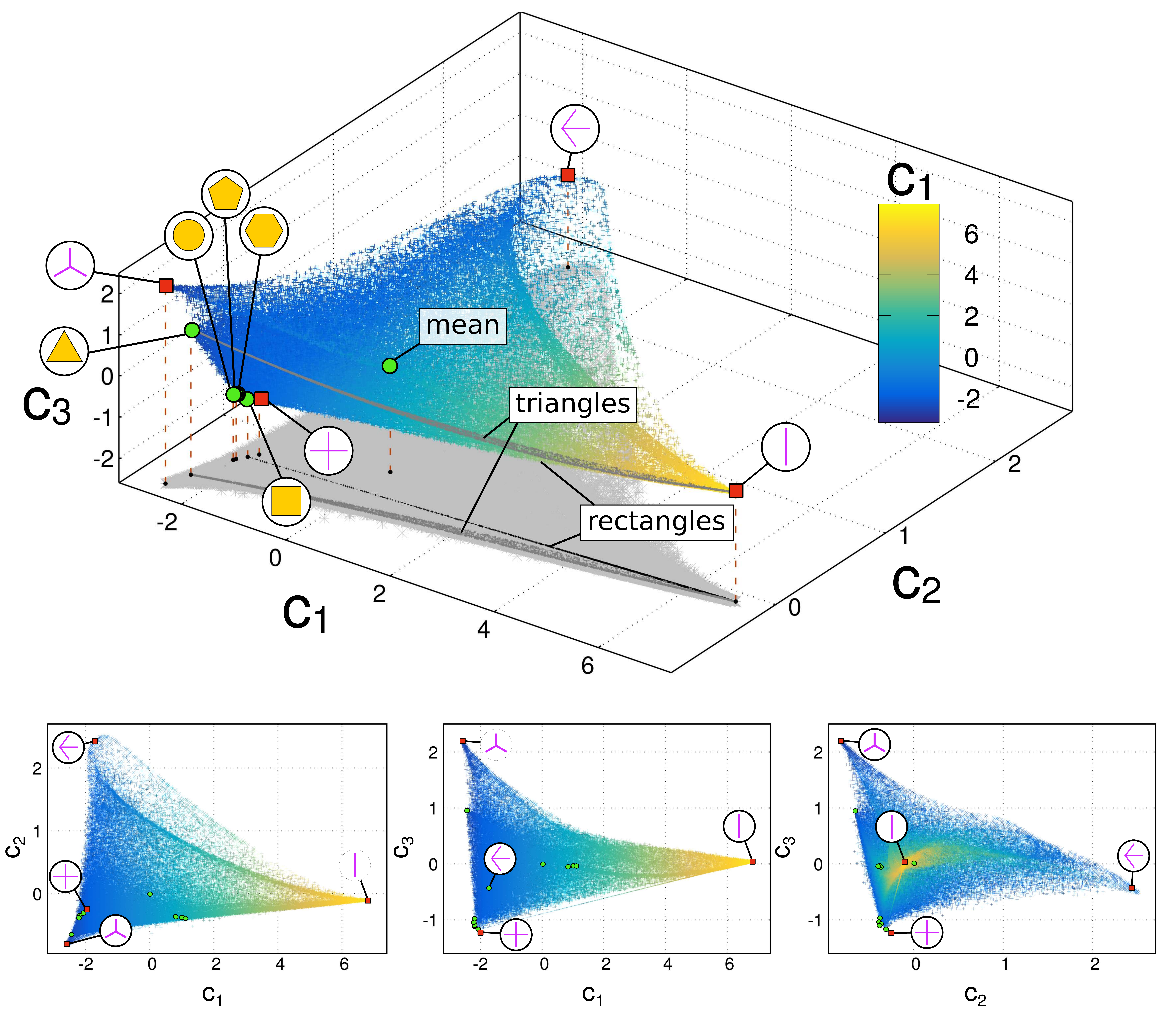}
\caption{Excluded areas in the space of the first three principal components. Top: The set of all values of the excluded area $A_{\text{exc}}(\phi)$
represented with the first three principal components $c_i$, $i=1,2,3$.
Each point represents one excluded area of the data set and is therefore associated to one particle shape.
The color code indicates the value of the first principal component $c_1$.
Green circles show the values of the excluded areas of selected particles, as indicated. Red squares indicate the four limiting cases 
analysed in the text. In light gray a projection onto the $c_1-c_2$ plane is shown.
In dark gray a projection of the excluded areas of triangles and rectangles is shown.
Bottom: Two dimensional projections of the excluded areas onto the $c_1-c_2$ (left), $c_1-c_3$ (middle), and $c_2-c_3$ planes.}
\label{fig5}
\end{figure*}

In Fig. \ref{fig5} we present the first three principal components $c_i$, $i=1,2,3$ of the excluded area of all particles in the set. Each point
represents the excluded area $A_{\text{exc}}(\phi)$ of one particle of the set. We also show two-dimensional projections onto the $c_1-c_2$, $c_1-c_3$, and $c_2-c_3$ planes.
The excluded areas are contained in a simply connected region with no holes and a well-defined boundary with four prominent limiting cases (highlighted
by red squares in Fig.~\ref{fig5}). We discuss in what follows these special limiting excluded areas found with PCA, together with their associated particle shapes.

{\bf First limiting case}. As indicated by the eigenvalues, the first principal component $c_1$ has the by far highest variance with values between $-3$ and $7.5$.
One vertex of the 3D projection of the excluded area, Fig.~\ref{fig5}, corresponds to the line segment, which is the limiting case
for which $c_1$ is maximized. In general, the value of the $c_1$ component increases with the elongation of the particles.
According to PCA, the elongation of a particle is therefore the most important geometric feature influencing the excluded area.
The excluded areas of the particles near this limiting case posses two well-defined minima, like in the case of a line segment shown in Fig.~\ref{fig2}a.

\begin{figure} 
 \includegraphics[width=1\columnwidth,angle=0]{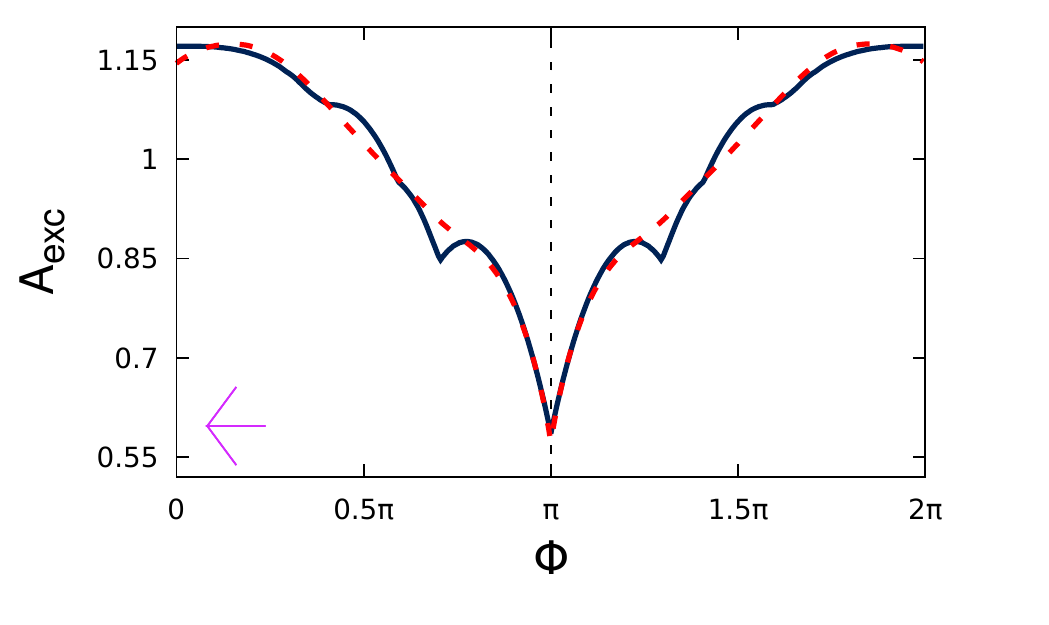}
 \caption{Excluded area (solid black line) and reconstructed excluded area using the first three principal components
 (dashed red line) as a function of the relative orientation $\phi$ for the particle presented in magenta (left bottom corner).}
 \label{fig6}
 \end{figure}

{\bf Second limiting case}. Maximizing the second principal component $c_2$ is another limiting case of the 3D projection,
see Fig.~\ref{fig5}. An illustrative excluded area and its corresponding particle shape are presented in Fig. \ref{fig6}.
The excluded areas in this region are characterized by a pronounced global minimum located at $\phi=\pi$, and a
global maximum near $\phi=0$. In Fig. \ref{fig6} both the actual excluded area and the reconstruction
using only the first three principal components are shown. The real excluded area has
secondary minima that are not captured by the reconstructed excluded area. However,
the overall agreement is very good and it justifies the use of only three principal components.
The secondary features that are not reproduced by the reconstructed excluded area might play a role
in the determination of the structure of phases with positional order but it is unlikely that they will affect the relative
stability of fluid mesophases with only orientational order.

The particles in this region of the 3D projection are line stars with three arms. It might be possible to
eliminate the secondary minima using shapes with curved lines (similar to the symbol $\in$).
The shape of the excluded area $A_{\text{exc}}(\phi)$ suggests that the particles prefer a state where
the neighbouring particles are antiparallel.

{\bf Third limiting case}. For the third limiting case emphasized by PCA, 
both $c_1$ and $c_2$ are minimal. The excluded areas located in this region have
three pronounced equidistant minima. Before further discussing this case we first have a look at the
excluded areas for polygons with three vertices, i.e. triangles, which are closely related.

\begin{figure*} 
 \includegraphics[width=0.9\textwidth,angle=0]{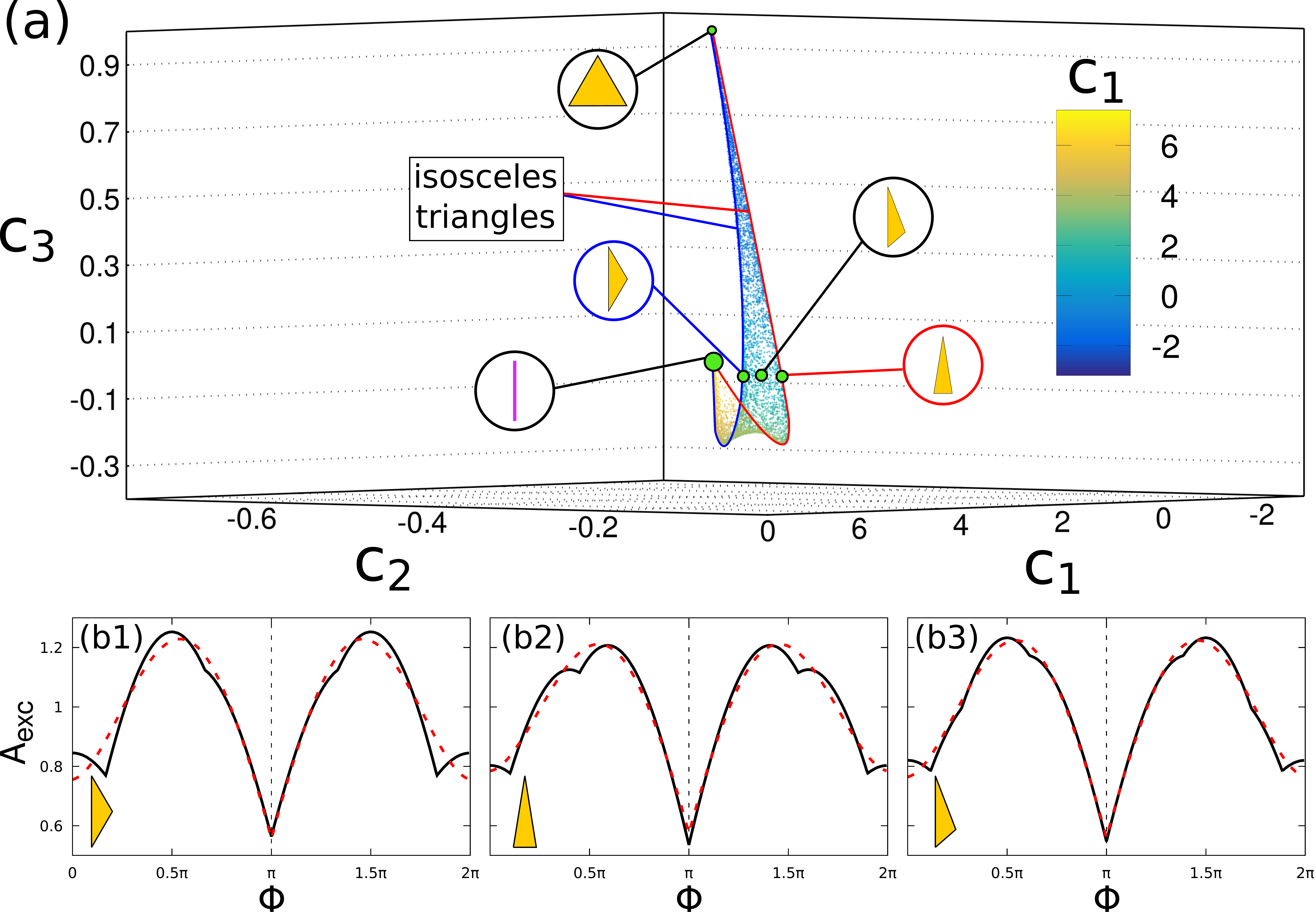}
 \caption{(a) Excluded areas of triangles represented with the first three principal components, $c_1,c_2$, and $c_3$.
 Each point represents the excluded area of one randomly generated triangle. The color indicates the value of the first principal component $c_1$. Green circles indicate
 the position in the space of principal components of the excluded areas corresponding to the depicted particle shapes.
 The blue (red) curve indicates the location of all obtuse (acute) isosceles triangles. 
 Bottom panels: Excluded areas (solid black line) and reconstructions using the first three principal components
 (dashed red line) as a function of the relative orientation $\phi$ for three particular triangles 
 (represented in the left bottom corner of each panel).}
 \label{fig7}
 \end{figure*}

The excluded areas of triangles in the base of the first three principal axes are highlighted in
dark grey in Fig.~\ref{fig5} and also represented in Fig. \ref{fig7}a.
All the excluded areas of triangles form a simply connected two-dimensional region which spans between
the two limiting cases given by a line segment and an equilateral triangle. These limiting cases correspond
to triangles with the maximum (line segment) and minimum (equilateral triangle) possible aspect ratios.
Going from the equilateral triangle to the line segment implies increasing the aspect ratio of a regular triangle.
There are two special ways of elongating an equilateral triangle: (i) taking one side and moving it away
from the opposite vertex, making the angle at the apex in the resulting acute isosceles triangle very small, and (ii)
taking one side and moving its two corresponding vertices away from each other along the direction of
this side, creating an obtuse isosceles triangle where the angle at the apex is very large. 
In both cases the intermediate triangles are isosceles and their excluded areas form the boundary of
the 3D projection in the base of principal components, see Fig.~\ref{fig7}a.
We have shown in Fig.~\ref{fig2}b the excluded area of an equilateral triangle, which has three global minima 
at $\{\pi/3,\pi,5\pi/3\}$ and three global maxima.
In Fig.~\ref{fig7}b we present the excluded area of other representative triangles.
In panel (b1) we show an obtuse isosceles triangle. The excluded area has only one global minimum at $\phi=\pi$.
The global minima of $A_{\text{exc}}$ at $\pi/3$ and $5\pi/3$ of an equilateral triangle are now local minima and have moved to 
a different relative orientation. 
In panel (b2) we represent the excluded area of an acute isosceles triangle. There is only one global minimum at $\phi=\pi$ and the position
of the secondary minima is also shifted with respect to that in an equilateral triangle.
In panel (b3) we present a non-isosceles triangle, which in the PCA analysis is located between the two previous cases (b1) and (b2), see Fig.~\ref{fig7}a.
The excluded area has characteristics of both acute and obtuse triangles.
In all three cases the reconstructions of the excluded areas neglect small features like the
the presence of local minima and kinks, but the overall agreement between the actual excluded area and that obtained with only three principal components is excellent.

 \begin{figure*}
 \includegraphics[width=0.9\textwidth,angle=0]{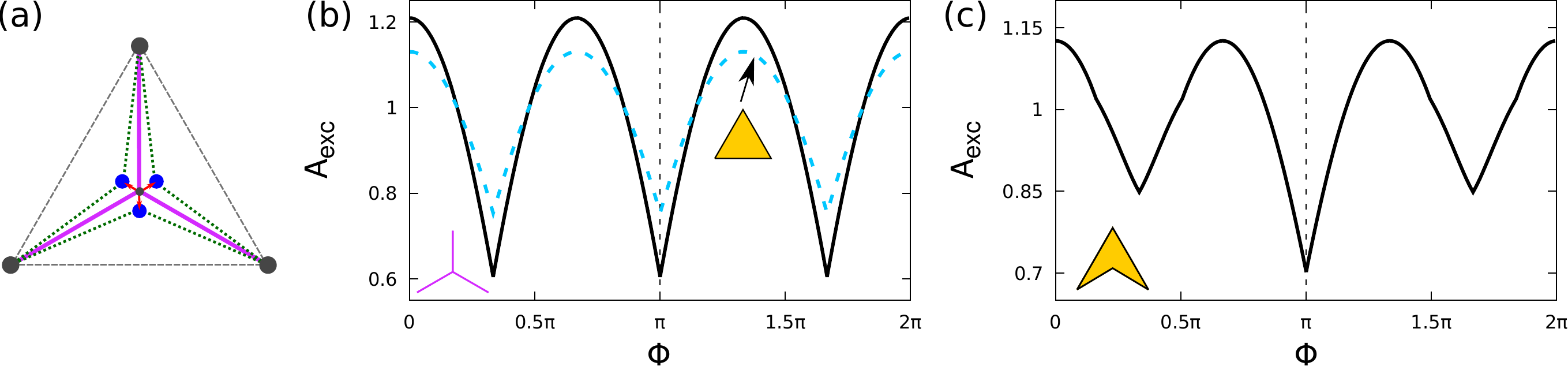}
 \caption{(a) Schematic of the third limiting case highlighted by PCA:
 a star line with three identical segments rotated by $2\pi/3$ (magenta) such that
 the outer vertices lie on top of an equilateral triangle (dashed grey). Splitting
 the inner vertex into three vertices (blue dots) creates a new particle shape (dotted green).
 (b) Excluded area for a star line with three identical segments (solid black) and an equilateral
 triangle (dashed blue) as a function of the relative orientation $\phi$. The star line is represented
 in magenta (bottom left corner). (c) Excluded area of an arrow-like particle, as shown
 in the bottom left corner.}
 \label{fig8}
 \end{figure*}

In the space of principal components, see Fig. \ref{fig5},  the excluded area of an equilateral triangle is located in the region where
both $c_1$ and $c_2$ are low, near the third limiting case for which $c_1$ and $c_2$ are minimal
and $c_3$ is maximal. At this location we find the excluded area of a star line particle made of three identical segments, any two
of them forming an angle of $2\pi/3$. A sketch of the particle is presented in Fig. \ref{fig8}a together with the
corresponding excluded area, Fig. \ref{fig8}b. The excluded area resembles that of an equilateral triangle
(also shown in the figure for comparison). Due to the normalization, Eq. \eqref{eq_norm}, the excluded area
of the star line appears to have a larger variance (difference between the maximum and the minimum values of the excluded area).
However, when normalizing the excluded area with the height of the particles both excluded areas have the same variance.
Strong differences are expected for three-body interactions and higher order terms.
Hence, a comparison between the bulk phase behaviour of this particle shape and that of equilateral triangles 
might help to understand the role of higher than two-body correlations on the bulk phase behaviour.

Another interesting property of this kind of particles is that the star line can be continuously deformed by splitting its
center and moving the resulting vertices radially towards the sides (see Fig. \ref{fig8}a). As a
special case, this includes particles where two of the inner vertices are located on top of the connecting
line between two of the outer vertices. The resulting particles have four vertices and the
shape of an arrow. The excluded area of such arrow-like particles is shown in Fig. \ref{fig8}c. The
main difference compared to the undeformed star particle is that the depth of two of the minima has
decreased. However, the minima are still located at the same orientations as in the initial star line.
This is in contrast to the case of triangles discussed above, for which any deformation simultaneously
changes the depth of the two secondary minima as well as the relative orientations at which they occur. 
In this case there is a complete family of particles in which the depth of the secondary minima can be tuned while
keeping their location (relative orientation) fixed. Particles of this type could present an isotropic-triatic
transition by increasing the density and then a second transition towards an uniaxial state at even higher densities.

{\bf Fourth limiting case}. The last limiting excluded area according to PCA, see \ref{fig5}, occurs
when $c_3$ is minimal. In this region a number of special particles is located. Among them we find disks, which
have a completely flat excluded area independent of the relative orientation.
Regular pentagons and hexagons together with other regular polygons with more
vertices are located near the disk. This can be explained with the decreasing
variance of the excluded area of regular polygons by increasing the number of sides, cf. Fig.~\ref{fig1}.

The regular polygon with the lowest value of $c_3$ is the square.
In the space of principal components, Fig.~\ref{fig5},
there is a continuous curve containing all possible
rectangles. At the end points of this curve we find the square and the line segment, which are the rectangles
with the minimal and maximal length-to-width aspect ratios, respectively.

\begin{figure}
  \includegraphics[width=0.9\columnwidth,angle=0]{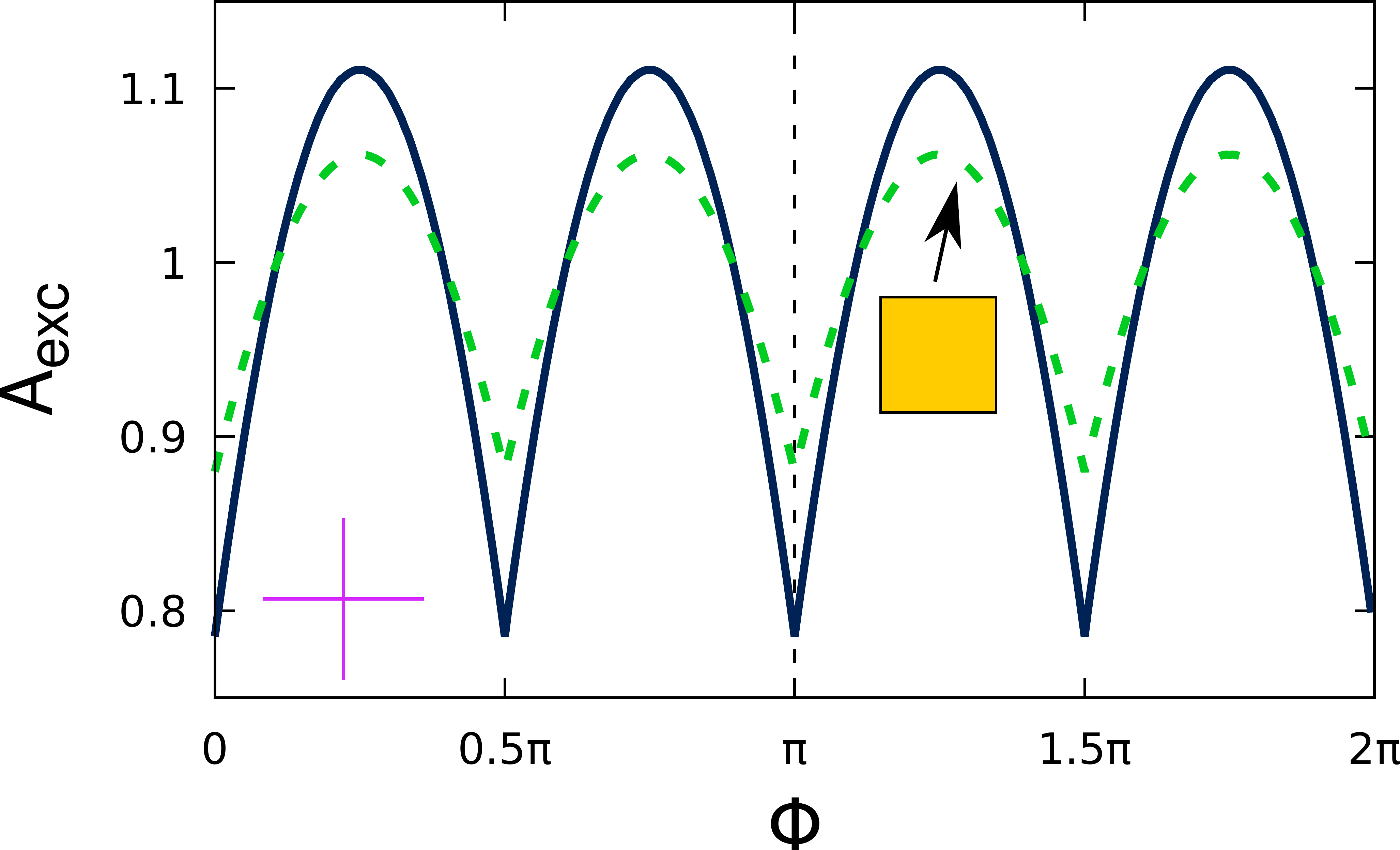}
  \caption{Excluded area for the fourth limiting case highlighted by PCA, 
  a star line with the shape of a plus (solid black line) and excluded area of a square (dashed green line)
  as a function of the relative orientation $\phi$. The star line particle is shown in the bottom left corner.}
  \label{fig9}
  \end{figure}

We have shown above two particle shapes, an equilateral triangle and a star line with three identical segments,
that share very similar excluded areas, see Fig.~\ref{fig8}a. A similar behaviour occurs for the case of a square and 
and a star line with four identical segments forming the shape of a plus, see Fig.~\ref{fig9}. 
The excluded area of the plus particle resembles that of the square but with higher variance
(see a comparison in Fig.~\ref{fig9}). The observed trend of decreasing variance in the
excluded area for regular polygons as the number of sides increases holds also for the case of
regular star lines.

The particle shapes of the four limiting excluded areas we have discussed above are well defined. This, however, is not the 
case for most excluded areas since different particle shapes can give rise to the same or almost the same excluded area.
That two different hard bodies can produce the same excluded volume has been recently proven for convex bodies~\cite{10.1088/1751-8121/aaf187}.

\subsection{Monte Carlo simulations}

We have shown above how PCA is useful to characterize the excluded areas of hard bodies (which play the role of the pair interaction
potential in soft systems). The next natural step is a complete analysis of the bulk phase behaviour of those particle shapes highlighted
by PCA using e.g., computer simulations or density functional theory. Some of the relevant shapes we have found,
like rectangles~\cite{doi:10.1063/1.1849159,Donev2006,thomas} and triangles~\cite{Gantapara2015, Martinez-Raton2018}
have been extensively studied, while others like the star lines have, to the best of our knowledge,
 not been analysed yet. Although such analysis is not the scope of this work, we have 
also performed short Monte Carlo simulations for selected particle shapes in order to have an initial understanding of how the particle shape
affects the bulk behaviour.

We performed Monte Carlo (MC) simulations in the $NpT$ (isothermal-isobaric) ensemble. In each simulation
a system of $N=200$ particles is compressed at constant pressure. The particles are randomly
initialized at sufficiently low density, so that the stable state is isotropic. 
Then $\sim5\cdot10^7$ Monte Carlo sweeps (MCS) are performed. Here, one MCS is an attempt to individually
move and rotate all particles in the system. After every $1.5\cdot10^4$ MCS, an attempt to slightly change
the volume of the system is performed. To this end, all particle positions are scaled accordingly. 
The maximum translation and rotation that each particle is allowed to perform in one MCS
as well as the maximum volume change in one step are chosen such that the total acceptance probability is approximately $0.25$.
We set a very high value of the pressure such that only steps that decrease the volume are accepted in order
to compress the system from the ideal gas limit to very high packing fractions.

\begin{figure*}
\includegraphics[width=0.9\textwidth,angle=0]{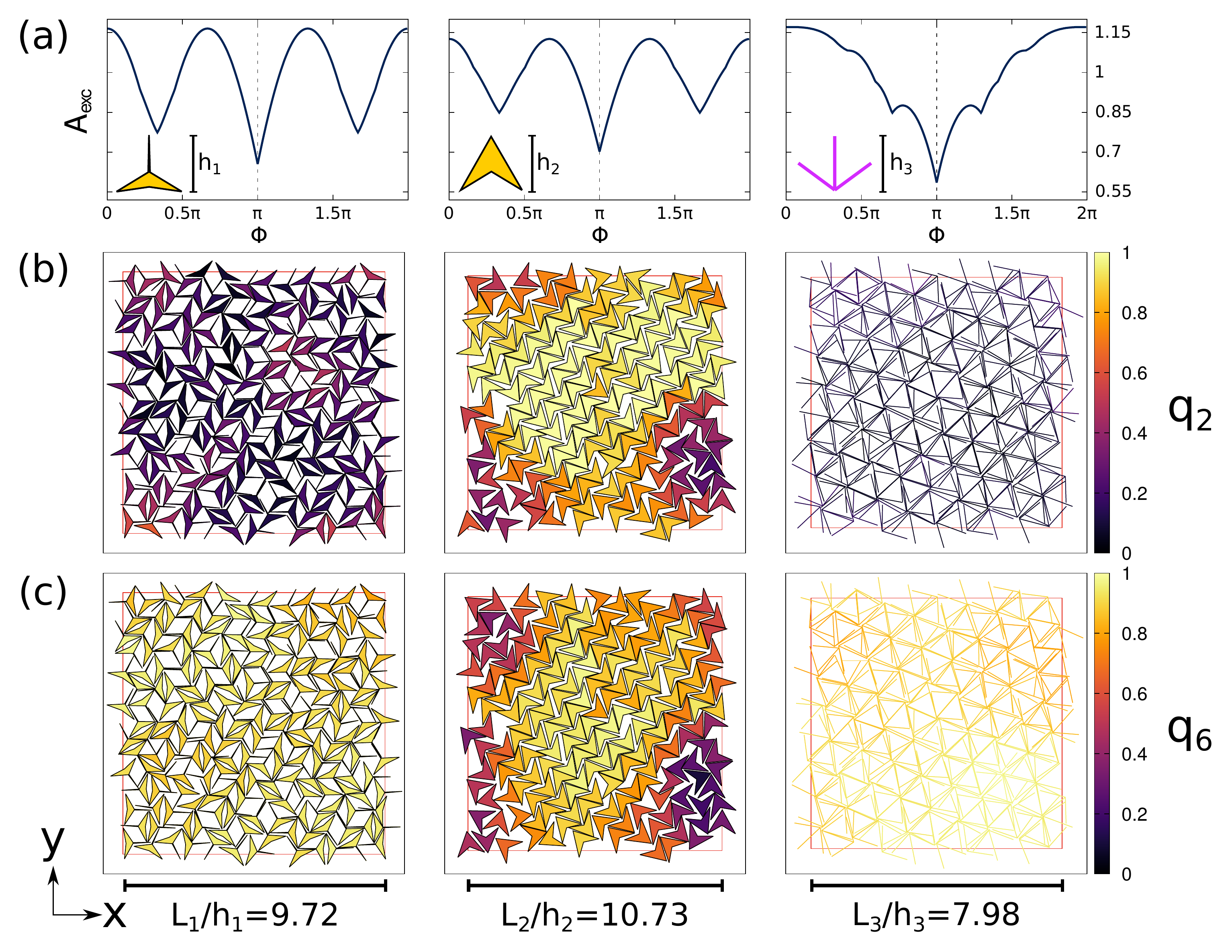}
\caption{(a) Excluded areas and geometry of three selected particles. Panels (b) and (c) show selected snapshots of a Monte Carlo simulation
of $N=200$ particles. The color code of the particles indicates the local particle-based order parameters $q_2$ (b) and $q_6$  (c).
The order parameters of each particle are computed taking into account all particles located in a circular region of diameter two particle
lengths centered at the particle of interest.}
\label{fig10}
\end{figure*}

We present here simulations for (i) particles that resemble an inverted umbrella (result of a deformation
of a star line with three segments as indicated in Fig.~\ref{fig8}a), (ii)
arrow particles, and (iii) the polar particles we showed in Fig. \ref{fig6}.
The shapes of the selected particles together with their excluded areas
are shown in panel (a) of Fig. \ref{fig10}.
Representative snapshots of particle configurations at high density are presented
in panels (b) and (c). The particles $(i=1,...,N)$ are colored according to their $q^{(i)}_2$ (b) and $q^{(i)}_6$ (c) orientational order parameters,
defined as
\begin{equation}
q^{(i)}_{k}=\left |\frac{1}{N_l} \sum_{j=1}^{N_l} e^{-ik\theta_{j}} \right|,\quad k=\{2,6\},
\end{equation}
where $\theta_{j}$ is the orientation of particle $j$, and the sum runs over the
$N_l$ particles located at a distance smaller than approximately two particle lengths from particle $i$ (including the $i-$th particle).

The orientational order of the inverted umbrella particles is triatic (three directors forming an angle of $\pi/3$ between them),
as indicated by the low $q_2$ (Fig.~\ref{fig10}b left) and high $q_6$ (Fig.~\ref{fig10}c left) values. In contrast, the orientational order
of the arrow-like particles is uniaxial (particles oriented on average along one direction), with high values of 
both $q_2$ (Fig.~\ref{fig10}b middle) and $q_6$ (Fig.~\ref{fig10}c middle). 
The different behaviour between these two particle shapes can be explained with the properties of the excluded area. The excluded areas of
both particles have the global minimum at $\phi=\pi$ and two local minima at $\phi=\pi/3$ and $5\pi/3$, see Fig.~\ref{fig10}a. The difference
lies in the ratio between the depths of the local minima and the global minimum (measured from the global maximum), which is $0.77$ for the inverted umbrella
(Fig.~\ref{fig10}a left) and $0.65$ for the arrow-like particle (Fig.~\ref{fig10}a middle).

The stable orientational order is the result of a competition between orientational and configurational entropies. In a triatic phase
the orientational entropy is maximized by populating the three possible minima of the excluded area, whereas in a uniaxial phase
the configurational entropy is maximized by populating only the global minimum (for which the packing is more efficient).
The result of this competition depends on the relative depth between the local and the global minima. This parameter can be continuously
tuned via deformation of the particle shape.

Note also that there is no minimum at $\phi=0$ for the excluded area of the arrow particles.
Therefore, in the uniaxial phase of the arrow particles neighbouring particles always point in opposite directions.
Our simulations suggest a possible coexistence between this ordered state and an isotropic state for the arrow particles.
Note that a small region of the simulation box remains isotropic. However, further longer and larger simulations are
required to study this in detail.\\

For the third type of particles (Fig.~\ref{fig10}a right) the excluded area has a well pronounced minimum at $\phi=\pi$.
According to this observation one could expect an uniaxial state similar to that found for the arrow particles.
The simulations, however, reveal that the particles prefer a state with triatic order. 
It is then clear that higher than two-body correlations are required to predict the bulk phase behaviour for this particle shape. 
This resembles the case of hard rectangles. There, particles with length-to-width aspect ratio smaller than $\sim7$ form small clusters
of a few particles each. The cluster formation leads to a global tetratic order~\cite{Velasco1,Velasco2,paper26} with two perpendicular
directors in which the symmetry differs from that of the particle shape.

\section{Discussion and Conclusions}

Using principal component analysis we have systematically
investigated the shape of the magnitude of the excluded area 
as a function of the particle orientation for two-dimensional hard-bodies. 
We have restricted the analysis to particle shapes given as
non-self-intersecting polygons and star lines. In both cases the particle shapes were randomly generated.
Despite the vast diversity of particle shapes, our analysis shows
that the variety of possible excluded areas is more restricted since global features dominate
the excluded areas. A clear indication of this is that only three principal components are
required to produce an excellent approximation of the magnitude of the excluded area
as a function of the relative particle orientation. 

One feature shared by
all excluded areas is the position of the global minimum which in our $\sim10^6$
randomly generated samples occurs always when the two particles are antiparallel, forming
a relative angle $\phi=\pi$. This result, known for convex bodies~\cite{Palffy-Muhoray2014}, seems to hold
also for non-convex particles.

The dimension reduction via PCA identifies the elongation of the particles as the most prominent
feature affecting the excluded area. Furthermore, when represented with the first three
principal components, all excluded areas are located on a simply connected three-dimensional region with no holes. 
At the boundaries we find four well-defined limiting cases corresponding to shapes of the excluded
area with one to four equidistant prominent minima. The relative depth between these prominent minima
is a parameter that can be adjusted by varying the particle shape. In contrast to this flexibility, there is not much freedom
to vary the relative angle at which the prominent minima of the excluded area occur.
However, other features of the excluded area like the position and number of secondary minima can be tuned.
While we do not expect a high impact on the fluid phases, the secondary features might play an important role on
the stability of phases with positional order.

We have identified different particle shapes with very similar excluded areas. These particles are ideal candidates
to investigate the role of higher than two-body correlations on the bulk phase behaviour. Higher than two-body
correlations can even dominate the behaviour of the system. We have shown an example using MC simulation in which the excluded
area possesses a unique global minimum at $\phi=\pi$ but the orientational order of the particles is triatic.

The phase behaviour of three particle shapes highlighted in our PCA analysis has been studied:
line segments form uniaxial nematic phases~\cite{Frenkel1985}, a fluid of squares~\cite{WOJCIECHOWSKI2004} or rectangles with short aspect ratio
form a tetratic phase~\cite{doi:10.1063/1.1849159}, and triangles form triatic phases~\cite{Gantapara2015}.
That is, the particle shapes highlighted in the PCA analysis give rise to the formation of mesophases
with different orientational properties. PCA is therefore a powerful technique to classify the interaction
in hard models and to anticipate the particle shapes of potential interest. PCA can complement other approaches intended
to understand self-assembly in colloidal systems such as the inverse design of pair potentials~\cite{doi:10.1063/1.5063802},
and the systematic study of regular shapes using computer simulations~\cite{C8SM01573B,2019}.
Recently it has been shown how PCA can be applied to detect phase transitions in lattice ~\cite{Wetzel2017} as well as in continuous~\cite{doi:10.1063/1.5049849} systems.

Promising extensions of the current work are the application of PCA to the excluded area of three-dimensional hard bodies and binary mixtures.
Regarding binary mixtures, we expect a much richer variety of shapes. Note for example that the excluded area between
two different particles does not have to be symmetric with respect to a certain relative orientation
(in contrast to the excluded area between identical particles, which is always symmetric with
respect to $\phi=\pi$). 
\begin{acknowledgments}
We thank M. Schmidt, E. Velasco, and Y. Mart{\'\i}nez-Rat{\'o}n for feedback and stimulating discussions.
\end{acknowledgments}

\end{document}